\documentclass[reprint, superscriptaddress, secnumarabic, amssymb, nobibnotes, aps, prl]{revtex4-1}

\usepackage{graphicx}
\usepackage{epstopdf}
\usepackage[T1]{fontenc}
\usepackage[latin9]{inputenc}
\usepackage{amsbsy}
\usepackage{gensymb}
\usepackage[T1]{fontenc}
\usepackage[latin9]{inputenc}
\usepackage{amsmath}
\usepackage{amssymb}
\usepackage{bbm}
\usepackage{braket}
\usepackage{xcolor}
\allowdisplaybreaks
\usepackage{graphicx}
\usepackage[colorlinks=true]{hyperref}  
\hypersetup{
    bookmarks=true,         
    unicode=false,          
    pdftoolbar=true,        
    pdfmenubar=true,        
    pdffitwindow=false,     
    pdfstartview={FitH},    
    pdftitle={},    
    pdfauthor={},     
    pdfsubject={},   
    pdfcreator={},   
    pdfproducer={}, 
    pdfkeywords={} {} {}, 
    pdfnewwindow=true,      
    colorlinks=true,       
    linkcolor=blue, 
    citecolor=blue,        
    filecolor=magenta,      
    urlcolor=blue           
} 
\usepackage[normalem]{ulem}

\newcommand{\equref}[1]{Eq.~(\ref{#1})}

\newcommand{\secref}[1]{Sec.~\ref{#1}}
\newcommand{\figref}[1]{Fig.~\ref{#1}}

\newcommand{\tableref}[1]{Table~\ref{#1}}

\begin{document}
\title{\textrm{Microscopic investigation of superconducting properties of a strongly coupled superconductor IrGe via $\mu$SR }}
\author{Arushi}
\affiliation{Department of Physics, Indian Institute of Science Education and Research Bhopal, Bhopal, 462066, India}
\author{Kapil~Motla}
\affiliation{Department of Physics, Indian Institute of Science Education and Research Bhopal, Bhopal, 462066, India}
\author{P. K. Meena}
\affiliation{Department of Physics, Indian Institute of Science Education and Research Bhopal, Bhopal, 462066, India}
\author{S. Sharma}
\affiliation{Department of Physics, Indian Institute of Science Education and Research Bhopal, Bhopal, 462066, India}
\author{D.~Singh}
\affiliation{ISIS Facility, STFC Rutherford Appleton Laboratory, Harwell Science and Innovation Campus, Oxfordshire, OX11 0QX, UK}
\author{P.~K.~Biswas}
\affiliation{ISIS Facility, STFC Rutherford Appleton Laboratory, Harwell Science and Innovation Campus, Oxfordshire, OX11 0QX, UK}
\author{A.~D.~Hillier}
\affiliation{ISIS Facility, STFC Rutherford Appleton Laboratory, Harwell Science and Innovation Campus, Oxfordshire, OX11 0QX, UK}
\author{R.~P.~Singh}
\email[]{rpsingh@iiserb.ac.in}
\affiliation{Department of Physics, Indian Institute of Science Education and Research Bhopal, Bhopal, 462066, India}
\email[]{rpsingh@iiserb.ac.in}

\date{\today}
\begin{abstract}
\begin{flushleft}
\end{flushleft}
Exploring superconductors which can possess pairing mechanism other than the BCS predicted s-wave have continually attracted considerable interest. Superconductors with low-lying phonons may exhibit unconventional superconductivity as the coupling of electrons with these low-lying phonons can potentially affect the nature of the superconducting ground state, resulting in strongly coupled superconductivity. In this work, by using magnetization, AC transport, specific heat, and muon spin rotation/relaxation ($\mu$SR) measurements, we report a detailed investigation on the superconducting ground state of the strongly coupled superconductor, IrGe, that has a transition temperature, T$_{C}$, at 4.7 K. Specific heat (SH), and transverse field $\mu$SR is best described with an isotropic s-wave model with strong electron-phonon coupling, indicated by the values of both $\Delta(0)/k_{B}T_{C}$ = 2.3, 2.1 (SH, $\mu$SR), and $\Delta C_{el}/\gamma_{n}T_{C}$ = 2.7. Zero-field $\mu$SR measurements confirm the presence of time-reversal symmetry in the superconducting state of IrGe.

\end{abstract}
\maketitle
\section{Introduction}

Understanding the pairing mechanism of unconventional superconductors is an active area of ongoing research in the field of superconductivity. In contrast to conventional superconductors, where weakly coupled electron-phonon superconductivity in spin-singlet configuration is proposed, in unconventional superconductors, the proposed pairing mechanisms strongly depends on the observed exotic features together with other symmetries present in them and cannot be generalized to all \cite{PM_SF1,PM_SF2,PM_NCS1,PM_NCS2,PM_NCS3,PM_NCS4,PM_p1,PM_p2,PM_LNG2}. Superconductors with low-lying phonons emerged as a class of unconventional superconductors with a very strongly coupled superconductivity \cite{SCS1,SCS2,SCS3,SCS4,SCS5,SCS6,SCS7,KOs2O6}. In these superconductors, the low-lying phonons originated from the rattling modes are proposed and also suggested that electron-rattler phonon mode coupling mediates the Cooper pairing, resulting in the extremely strong-coupled superconductivity. These excitations may behave differently from the usual phonons, which give rise to conventional BCS superconductivity and may have an impact on the order parameter symmetry. Only a few superconductors with low-lying phonons resulting in a strong coupling have been studied via microscopic and macroscopic measurements. It provides the presence of an isotropic energy gap in RbOs$_{2}$O$_{6}$ \cite{RbOs2O6_MuSR}, anisotropic or multigap with small energy in KOs$_{2}$O$_{6}$ \cite{KOs2O6_MuSR}, multiband superconductivity in SrPt$_{3}$P \cite{SrPt3P_MuSR} and  Nb$_{5}$Ir$_{3}$O \cite{Nb5Ir3O}. The existence of exotic superconducting gap structures and strong electron-phonon coupling in these compounds may hint towards the role of low-lying phonons and requires a schematic study of more compounds that fall into this category to completely understand the role of low-lying phonons on the nature of superconducting ground state properties.

The transition metal germanide, IrGe, is also known to exhibit strongly coupled superconductivity with a high value of specific heat jump (C$_{el}/\gamma_{n}T_{C}$ = 3.04) and superconducting gap ($\Delta(0)/k_{B}T_{C}$ = 2.57) \cite{IrGe}. The Debye temperature from the specific heat measurements gives, $\theta_{D}$ = 160 K, which is almost half of the similar molar mass compound, PtGe. Phonon density of state calculations suggests the presence of low-lying phonons in IrGe \cite{IrGe_SOC}. The obtained Einstein temperature from specific heat measurement account for the energy of low-lying phonons, which falls in the temperature range calculated for the rattling modes in superconducting pyrochlore osmates \cite{IrGe} where these modes are considered as the origin of low-lying phonons \cite{KOs2O6_MuSR}. Moreover, the density of states at the Fermi level is mainly dominated by the Ir 5d states \cite{IrGe_SOC} which can induce strong SOC effects and also the emergence of 5d superconductivity. Therefore, the correlation between unconventional superconductivity and strong SOC, and to examine the influence of low lying excitations on the nature of superconducting ground state warrants a thorough investigation on IrGe through a microscopic tool such as $\mu$SR. This study will also serve as a reference for the strongly coupled low lying phonon mediated superconductors as a detailed study on any of these compounds has not been performed till now. 

In this paper, we report the superconducting ground state properties of IrGe through different techniques such as AC transport, magnetization, specific heat and $\mu$SR. Bulk superconductivity was confirmed from all the mentioned measurements at T$_{C}$ = 4.74(3) K. Transverse field $\mu$SR measurements together with specific heat reveals a nodeless s-wave superconducting gap with strong electron-phonon coupling. Zero-field $\mu$SR measurements confirm the presence of time-reversal symmetry in IrGe.  

\section{Experimental Details}
\vspace{-10pt}
The polycrystalline sample of IrGe was prepared by the solid-state reaction method described in the Ref. \cite{IrGe}. Powder x-ray diffraction (XRD) experiment was conducted using a PANalytical X'Pert Pro diffractometer equipped with CuK$_{\alpha}$ radiation ($\lambda$ = 1.5406 \text{\AA}).  A Quantum Design Superconducting Quantum Interference Device (SQUID) was used to collect magnetization data in the temperature range of 1.8 - 7.0 K with various applied fields. Specific heat measurements and AC transport without and with fields were performed on a Quantum Design Physical Property Measurement System (PPMS). Specific heat measurements employed a two-tau relaxation method whereas for AC transport, a four probe method was used. The $\mu$SR measurements were conducted at the ISIS Neutron and Muon Facility, STFC Rutherford Appleton Laboratory, United Kingdom using the MUSR spectrometers in both the longitudinal and transverse directions \cite{Muon}. The powdered sample of IrGe was mounted on a silver (99.995\%) holder using GE varnish and inserted in the sample chamber of a He-3 sorption cryostat. 100 \% spin-polarized muons were implanted into the sample which decays into positrons and neutrinos after a mean lifetime of 2.2 $\mu$s. The decayed positrons are emitted preferentially in the direction related to the orientation of the muon spin vector and were collected by either forward (F) or backward (B) detectors in longitudinal configuration. The time-dependent asymmetry G$_{z}$(t) is the measured quantity and is calculated as (N$_{F}$(t)-$\alpha$N$_{B}$(t))/(N$_{F}$(t)+$\alpha$N$_{B}$(t)), where N$_{F}$(t) and N$_{B}$(t) are the number of positron counts in the forward and backward detector, respectively. $\alpha$ is an instrumental calibration factor which represents the efficiency for relative counting between F and B detectors. Thus, the time evolution of muon polarization (calculated from G$_{z}$(t)) allows one to determine the local magnetic environment experienced by the muon ensemble. Zero-field (ZF) measurements were carried out with detectors placed in longitudinal configuration. To achieve true zero fields, stray magnetic fields (originated from Earth and neighbouring instruments) were cancelled within 1 $\mu$T limit by a fluxgate magnetometer and an active compensation system controlling three pairs of correction coils. In transverse field (TF) measurements, a field perpendicular to the initial muon spin polarization direction was applied. TF measurements for IrGe were performed in the vortex state by applying a field of 50 mT in field cooled condition. 

\begin{figure}
\includegraphics[width=1.05\columnwidth, origin=b]{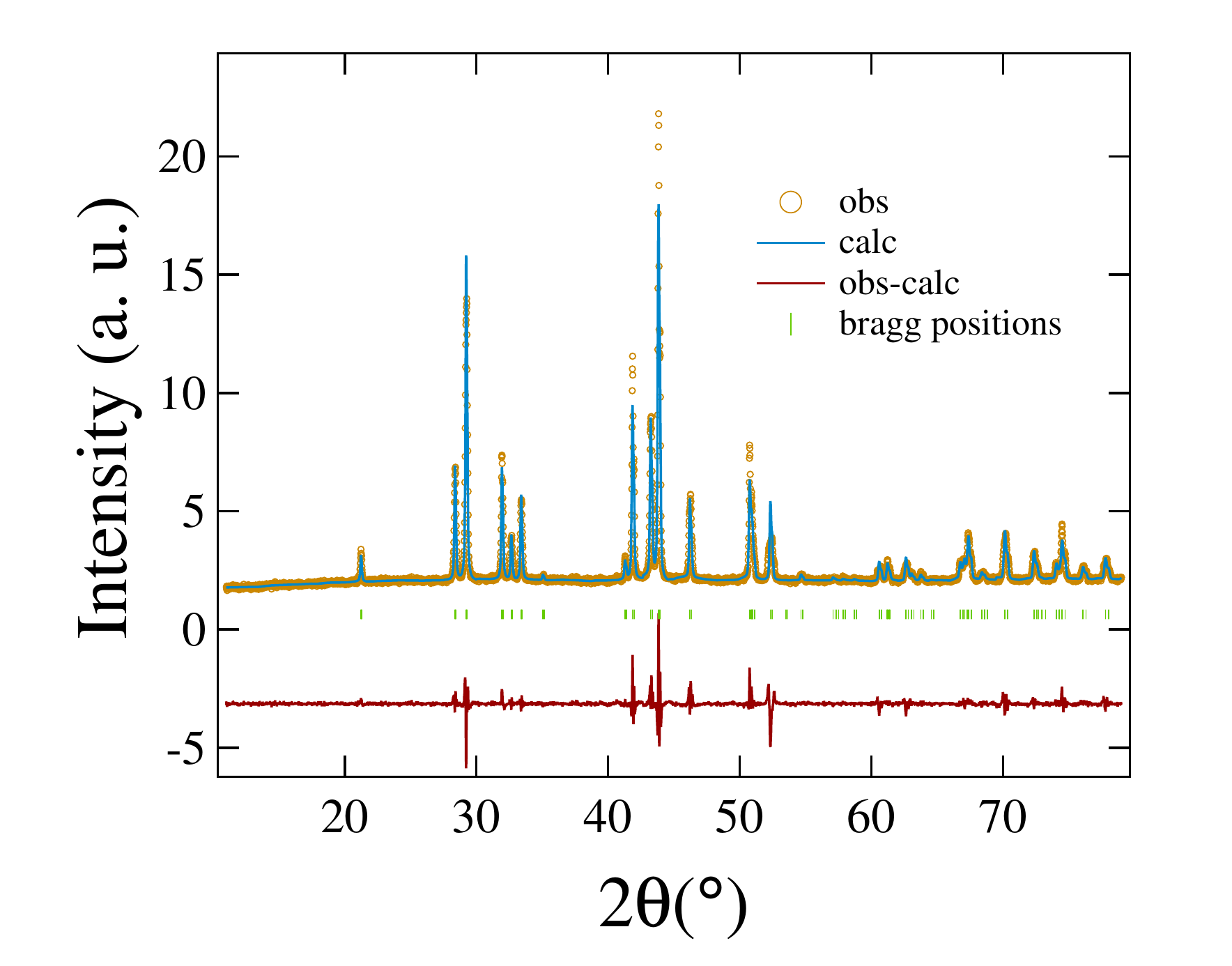}
\caption{\label{Fig1:XRD} a) Room temperature powder XRD pattern of IrGe is shown by yellow circles whereas blue line represents the Rietveld refinement. Bragg positions are shown by green vertical bars and the red solid line displayed the difference between calculated and observed pattern.}
\end{figure}

\section{Results and Discussion}

\subsection{Sample Characterization} 
\vspace{-7pt}
Rietveld refinement is shown in \figref{Fig1:XRD} confirmed that material crystallizes in orthorhombic structure with space group: P$nma$ \cite{IrGe}. The lattice parameters and atomic positions obtained from the refinement are given in \tableref{Tab:XRD}.

\begin{table}[h!]
\caption{Parameters obtained from Rietveld refinement of IrGe}
\label{Tab:XRD}
\begin{tabular}{c r} \hline\hline
Structure& Orthorhombic\\
Space group&        P$nma$\\ [1ex]
Lattice parameters\\ \hline
a (\text{\AA})&  5.6000(1)\\
b (\text{\AA})&  3.4935(1)\\
c (\text{\AA})&  6.2865(1)\\

\end{tabular}
\\[1ex]
\begingroup
\setlength{\tabcolsep}{4pt}
\begin{tabular}[b]{c c c c c c}
Atom&  Wyckoff position& x& y& z& \\[1ex]
\hline
Ir& 4c& 0.00141& 0.25000& 0.20324&\\ 
Ge& 4c& 0.18524& 0.25000& 0.58013& \\
[1ex]
\hline
\end{tabular}
\par\medskip\footnotesize
\endgroup
\end{table}

\begin{figure}
\includegraphics[width=1.05\columnwidth, origin=b]{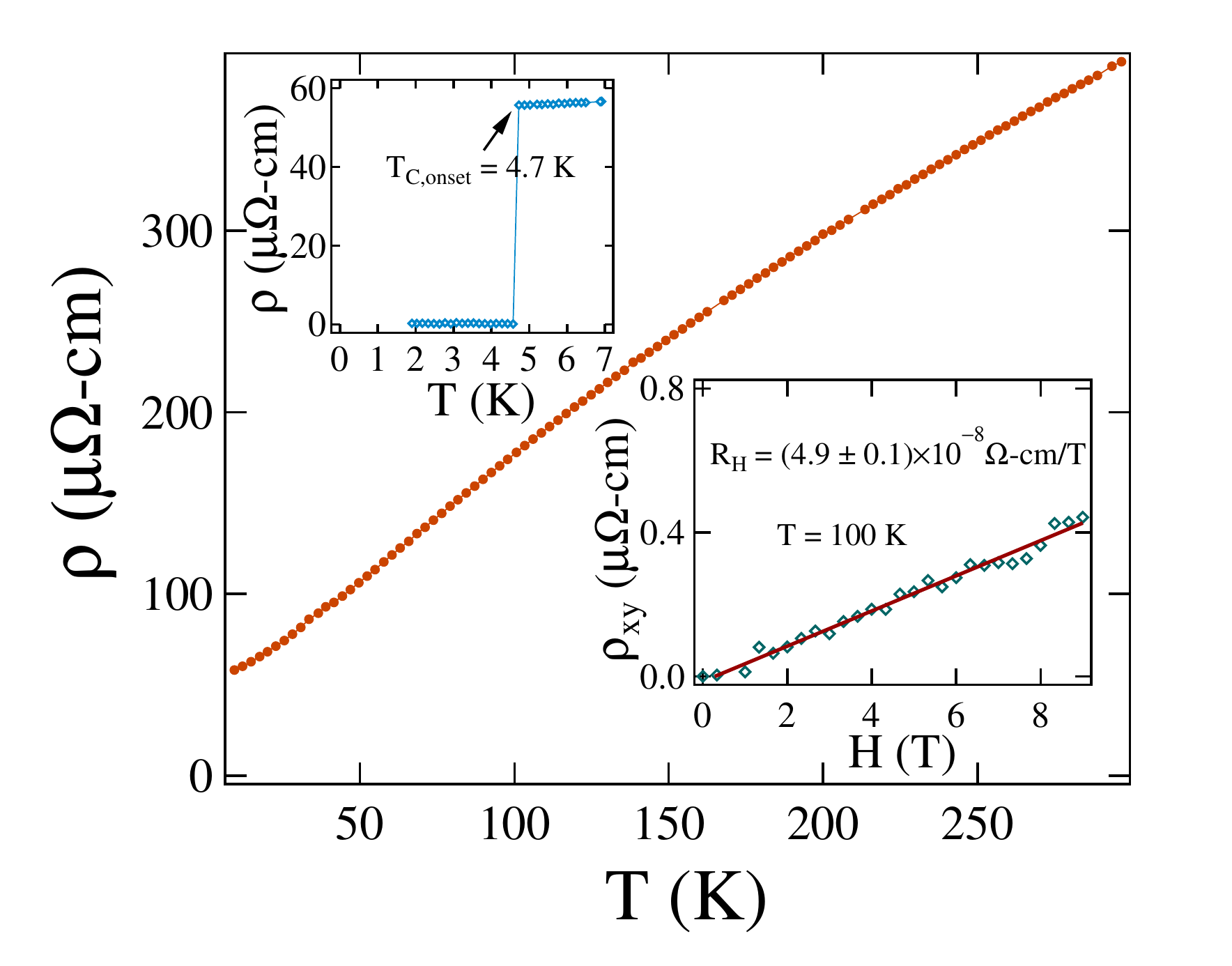}
\caption{\label{Fig2:Res} a) Temperature dependence of resistivity in the temperature range 10 K $\leq$ T $\leq$ 300 K. Inset in top left corner shows the resistivity drop at T$_{C,onset}$ = 4.7(1) K. Bottom right inset represents the field variation of hall resistivity up to + 9 T at T = 100 K.}
\end{figure}

\begin{figure*}
\includegraphics[width=2.0\columnwidth, origin=b]{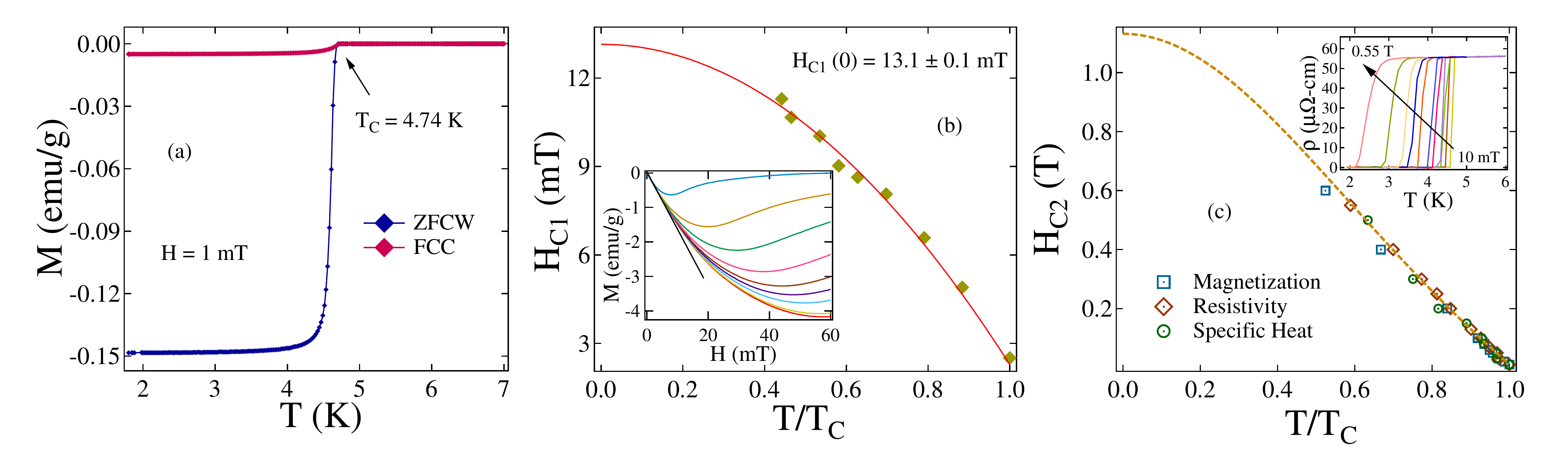}
\caption{\label{Fig3:Mag} a) The magnetic moment w.r.t. temperature measured at 1 mT. b) The lower critical field vs temperature where the inset shows the low-field magnetization curves at various temperatures. c) The temperature dependence of the upper critical field from various measurements where the inset shows the field and temperature variations of $\rho$.}
\end{figure*}

\subsection{Normal and superconducting state properties}
\vspace{-7pt}
The main panel of \figref{Fig2:Res} displays the temperature dependence of resistivity, $\rho(T)$ above the superconducting state and in zero applied field. Inset on the top left corner shows the enlarged view of $\rho(T)$ in the low-temperature regime where it exhibits a drop in resistivity value at T$_{C,onset}$ = 4.7(1) K and the zero resistance state is achieved at T$_{C,0}$ = 4.6(1) K with a width of superconducting transition, $\Delta$T = 0.1 K. The normal state $\rho(T)$ shows an almost linear T dependence down to low temperatures similar to a strongly coupled superconductor SrPt$_{3}$P \cite{SCS6} where such behaviour is attributed to the presence of low-lying excitations. At high temperatures, the resistivity curve seems to saturate which implies that the mean free path of electrons reaches the order of inter-atomic spacing. A similar response is observed for other strongly coupled A15 superconductors \cite{A15} and SrPt$_{3}$P \cite{SCS6} where the saturation of resistivity is interpreted as the strong coupling between the charge carriers and phonons. Hall measurement was also performed to calculate the carrier concentration and the type of charge carriers. Inset in the bottom right shows the field dependence of hall resistivity ($\rho_{xy}$) measured at T = 100 K. $\rho_{xy}$(H) is well described by a straight line fit and provides a slope which is the hall coefficient R$_{H}$ = 4.9(1)$\times$10$^{-8}$ $\Omega$-cm/T. The positive sign indicate the hole carrier concentration, and the relation R$_{H}$ = $1/ne$ yields $n = 1.27(2)\times10^{-28} m^{-3}$.

\begin{figure} 
\includegraphics[ width=1.0\columnwidth]{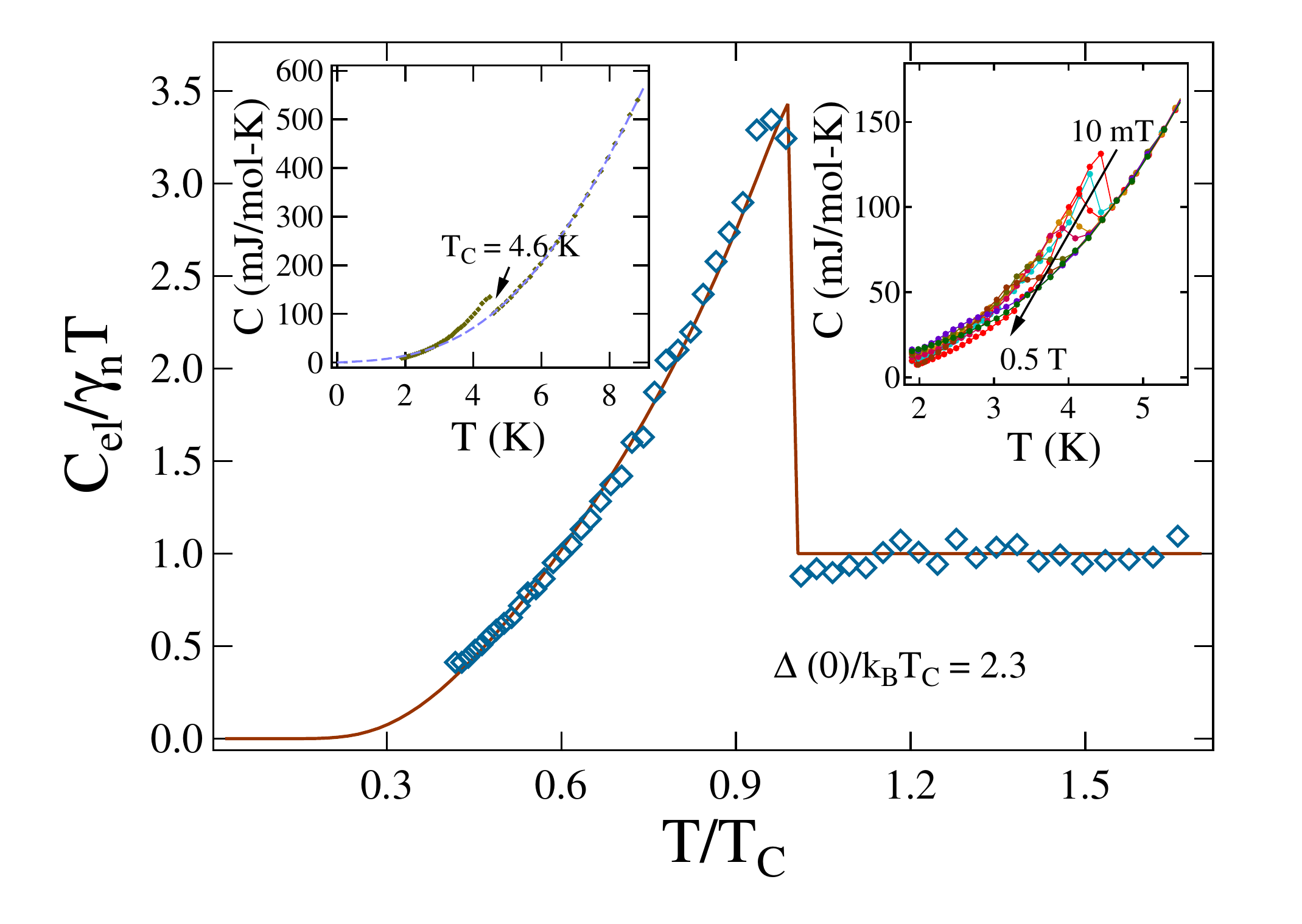}
\caption{\label{Fig4:SH} Temperature dependence of normalized specific heat in zero applied fields where solid red line is a fit using isotropic s-wave model. Inset(Left): Specific heat variation with T which exhibit a sharp jump at T$_{C}$ = 4.6 K. Dotted line is a fit using C = $\gamma_{n}$T + $\beta_{3}$T$^{3}$ to calculate the electronic and phononic contributions. Inset(Right): Specific heat w. r. t. temperature at various applied fields starting from 10 mT to 0.5 T.}
\end{figure}

Magnetic susceptibility of IrGe in an applied field of 1 mT via two different modes: i) zero-field cooled warming (ZFCW), and ii) field cooled cooling (FCC) is shown in \figref{Fig3:Mag}(a). It exhibits a clear diamagnetic signal below a transition temperature, T$_{C}$ = 4.74(3) K and confirms the bulk superconductivity. The difference in the magnitude of the FCC diamagnetic signal than the ZFCW can be accounted due to the strong flux pinning present in the material. To calculate the lower critical field, H$_{C1}$(0), low-field magnetization curves were obtained at different temperatures,  from 1.9 K to 4.3 K. H$_{C1}$ is defined as the point deviating from the initial slope of linear or Meissner line for individual temperature curves. The corresponding data sets are shown in \figref{Fig3:Mag} (b) and fitted with the Ginzburg-Landau relation given below and yield H$_{C1}(0)$ = 13.1(2) mT
\\
\begin{equation}
H_{C1}(T)=H_{C1}(0)\left[1-\left(\frac{T}{T_{C}}\right)^{2}\right]
\label{Hc1}
\end{equation} 
\\ 
The Upper critical field H$_{C2}$(0) is obtained from a range of measurements: magnetization, resistivity, and specific heat (see \figref{Fig4:SH} right Inset) at different applied fields. In magnetization, the onset of superconductivity is taken as the criteria for T$_{C}$, whereas in specific heat and resistivity, the midpoint is considered as the transition temperature. Figure \ref{Fig3:Mag}(c) represents H$_{C2}$(T) from all measurements, and it follows a linear behaviour near T = T$_{C}$ that can be fitted well with the phenomenological Ginzburg-Landau expression below where $t$ = T/T$_{C}$. The fit provides H$_{C2}$(0) = 1.13(2) T.
\\
\begin{equation}
H_{C2}(T)=H_{C2}(0)\frac{\left[1-t^{2}\right]}{\left[1+t^{2}\right]}
\label{Hc1}
\end{equation} 
\\

 There are two main mechanisms that can destroy superconductivity: a) orbital limiting effect, b) Pauli paramagnetic limit. When the kinetic energy of Cooper pairs exceeds the condensation energy, the orbital effect comes into play, whereas in the case of the Pauli limiting effect, the spin of one of electron in a Copper pair is oriented in the direction of the applied field, which in turn destroy superconductivity. The orbital limit of an upper critical field, H$_{C2}^{orb}$(0) for a type II dirty limit superconductor (see \secref{Elec}) is given by \cite{WHH1,WHH2} 
 \begin{equation}
 H_{C2}^{orb}(0) = -\alpha T_{C}\left.\frac{dH_{C2}(T)}{dT}\right|_{T=T_{C}}
 \end{equation}
 where $\alpha$ = 0.693 for dirty limit superconductors, T$_{C}$ = 4.7 K and the slope at T = T$_{C}$ is given by 0.21(3) T and yields $H_{C2}^{orb}(0)$ = 0.68(2) T. The Pauli limiting field is given by \cite{Pauli1,Pauli2} H$_{C2}^{P}$ = 1.86 T$_{C}$, and T$_{C}$ = 4.74 K. It provides H$_{C2}^{P}$ = 8.74(2) T. The Maki parameter describes the relative importance of both the orbital and paramagnetic effects in suppressing superconductivity and is given by \cite{Maki} $\alpha_{M} = \sqrt{2}H_{C2}^{orb}(0)/H_{C2}^{p}(0)$ = 0.11(2). The obtained value of $\alpha_{M}$ is significantly less than unity, suggesting that the influence of the Pauli limiting effect is small. The two fundamental length scales of any superconductor are the coherence length $\xi_{GL}(0)$, and penetration depth $\lambda_{GL}(0)$. These are calculated from H$_{C1}(0)$ and H$_{C2}(0)$. $\xi_{GL}(0)$ is 170(1) $\text{\AA}$ using the formula \cite{Xi_GL}: $H_{C2}(0) = \frac{\Phi_{0}}{2\pi\xi_{GL}^{2}}$ where $ \Phi_{0}$ ( = 2.07 $\times$ 10$^{-15}$ Tm$^{2}$) is the magnetic flux quantum and H$_{C2}$(0) = 1.13(2) T. The London penetration depth ($\lambda_{GL}$(0)) is calculated using the expression \cite{lambda_GL}: 
\\
\begin{equation}
H_{C1}(0) = \frac{\Phi_{0}}{4\pi\lambda_{GL}^2(0)}\left(\mathrm{ln}\frac{\lambda_{GL}(0)}{\xi_{GL}(0)}+0.12\right)   
\label{eqn6:ld}
\end{equation} 
\\
which takes into account both $\xi_{GL}(0)$ = 170(1) $\text{\AA}$ and H$_{C1}(0)$ = 13.1(2) mT, and provides $\lambda_{GL}(0)$ = 1750(11) $\text{\AA}$. The Ginzburg-Landau parameter is given by $\kappa_{GL} = \frac{\lambda_{GL}(0)}{\xi_{GL}(0)}$ = 10(1) $>$ 1/$\sqrt2$, classifies IrGe as a type II superconductor. The thermodynamic critical field H$_{C}(0)$ was estimated using the relation: $H_{C1}(0)H_{C2}(0) = H_{C}^2\mathrm{ln}\kappa_{GL}$ \cite{lambda_GL} which yields H$_{C}(0)$ as 78 mT.

Figure \ref{Fig4:SH} inset presents the temperature dependence of the specific heat at zero applied field. A sharp jump at T$_{C}$ = 4.6 K is evidence of  bulk superconductivity. The normal state data was fitted with the formula: C = $\gamma_{n}$T + $\beta_{3}$T$^{3}$ where $\gamma_{n}$ is the Sommerfeld constant of the normal state, and $\beta_{3}$ represents specific heat coefficient of the lattice part. The fit provides $\gamma_{n}$ = 3.1(2) mJ mol$^{-1}$K$^{-2}$, and $\beta_{3}$ = 0.94(1) mJ mol$^{-1}$K$^{-4}$. The Debye temperature can be written as $\theta_{D}$ = $({12\pi^{4}RN}/{5\beta_{3}})^{1/3}$ \cite{thetaD} where R = 8.314 J mol$^{-1}$K$^{-2}$ is a gas constant, and N = 2 is the number of atoms in the formula unit. Using the relation, we obtained $\theta_{D}$ = 160(1) K. McMillan \cite{McMillian} model is used to evaluate the strength of electron-phonon coupling, $\lambda_{e-ph}$ which is related to $\theta_{D}$ and T$_{C}$ as follows:  
\\
\begin{equation}
\lambda_{e-ph} = \frac{1.04+\mu^{*}\mathrm{ln}(\theta_{D}/1.45T_{C})}{(1-0.62\mu^{*})\mathrm{ln}(\theta_{D}/1.45T_{C})-1.04 }
\label{eqn8:ld}
\end{equation}
\\
Here $\mu^{*}$ represents the screened Coulomb potential and taken to be 0.13. The calculated value of $\lambda_{e-ph}$ = 0.78(2) indicates the strong electron-phonon coupling in the superconducting state of IrGe. To determine the electronic contribution of the specific heat, we have subtracted the phononic part ($\beta_{3}$T$^{3}$) from the total specific heat (C). Low temperature electronic specific heat data can tell us about the nature of the superconducting gap structure, and in this regard, the normalized electronic specific heat was fitted, and yields an excellent fit, with just a single isotropic s-wave gap expression which is related to the normalized entropy by
\\
\begin{equation}
\frac{C_{el}}{\gamma_{n}T_{C}} = t\frac{d(S/\gamma_{n}T_{C})}{dt} \\
\label{eqn8:BCS2}
\end{equation}
\\
where the BCS expression for the normalized entropy is provided below:
\\
\begin{equation}
\frac{S}{\gamma_{n}T_{C}} = -\frac{6}{\pi^2}\left(\frac{\Delta(0)}{k_{B}T_{C}}\right)\int_{0}^{\infty}[ \textit{f}\ln(f)+(1-f)\ln(1-f)]dy \\
\label{eqn7:BCS1}
\end{equation}
\\
where $\textit{f}$($\xi$) = [exp($\textit{E}$($\xi$)/$k_{B}T$)+1]$^{-1}$ is the Fermi function, $\textit{E}$($\xi$) = $\sqrt{\xi^{2}+\Delta^{2}(t)}$, where E($ \xi $) is the energy of the normal electrons measured relative to Fermi energy, $\textit{y}$ = $\xi/\Delta(0)$, $\mathit{t = T/T_{C}}$ and $\Delta(t)$ = tanh[1.82(1.018(($\mathit{1/t}$)-1))$^{0.51}$] is the BCS approximation for the temperature dependence of energy gap. The expression provides a good fit to the data shown by a red solid line in \figref{Fig4:SH} with the gap value $\Delta(0)$/k$_{B}$T$_{C}$ = 2.3(1). The value of the gap significantly exceeds that of the predicted BCS value, 1.76 and suggests that strongly coupled superconductivity is present in IrGe.

\begin{figure} 
\includegraphics[width=1.0\columnwidth, origin=b]{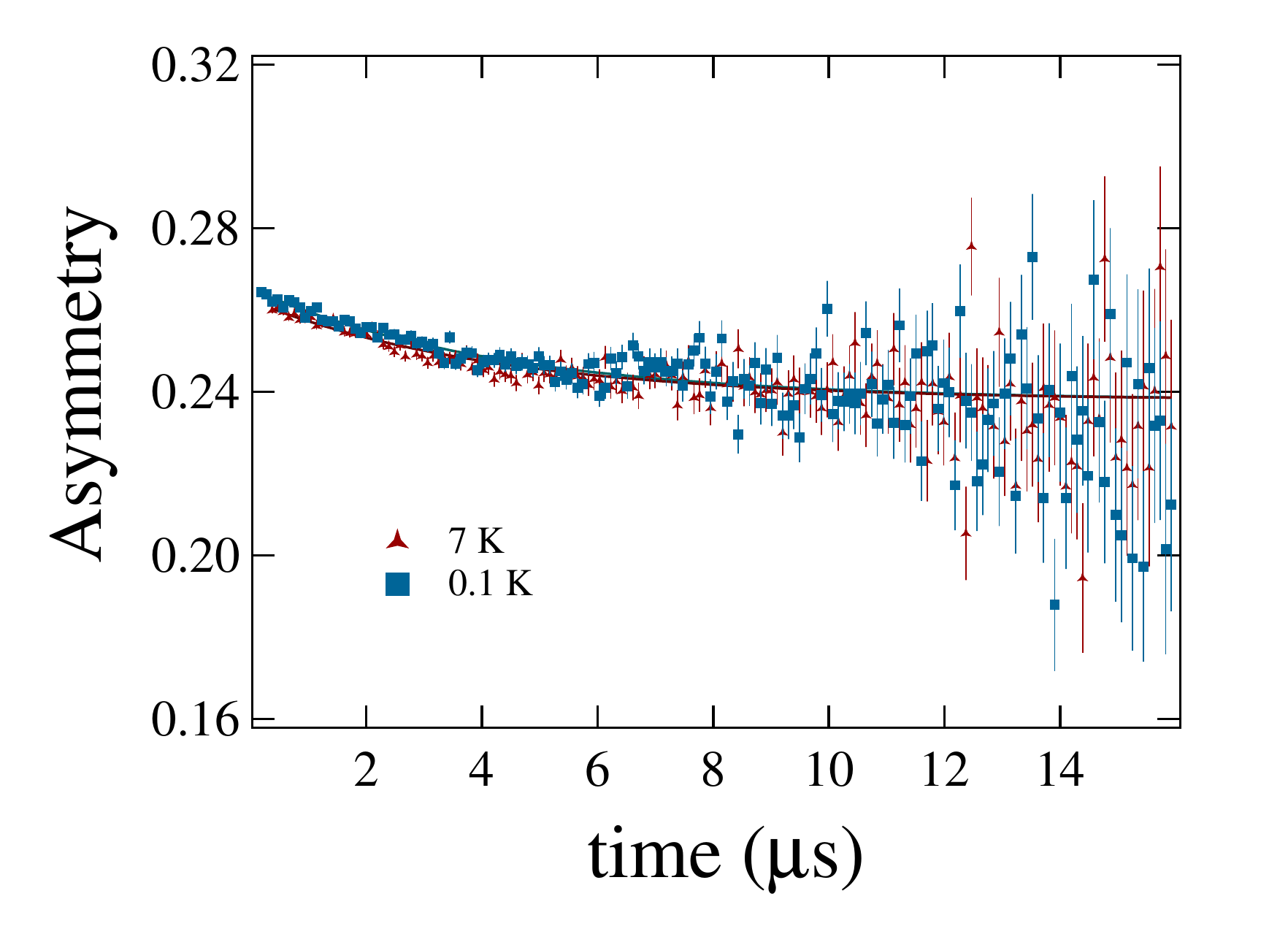}
\caption{\label{Fig5:ZF} Time-domain ZF-$\mu$SR spectra of IrGe in the superconducting (0.1 K) and normal state (7.0 K). The solid lines are fit using \equref{eqn1:ZF}. }
\end{figure}

\subsection{Muon spin rotation and relaxation}
\vspace{-7pt}
Zero-field $\mu$SR measurements were performed to confirm the presence of possible magnetism and/or breaking of time reversal symmetry in the superconducting state of IrGe. Figure \ref{Fig5:ZF} presents the time evolution of ZF $\mu$SR asymmetry spectra collected above and below T$_{C}$. For broken TRS, the onset of spontaneous magnetization causes an increase in the relaxation rate in the superconducting state \cite{RecentRebased}. However, for our case, there is no change in the relaxation rate across T$_{C}$ which confirms the absence of time-reversal symmetry breaking in IrGe. Both the spectra are well described with damped Kubo-Toyabe function plus a flat background term given below \cite{ZF}:
\begin{equation}
\begin{split}
A(t)=A_{0}[\frac{1}{3}+\frac{2}{3}(1-\sigma_{zf}^{2}t^{2})\exp\left(-\frac{1}{2}\sigma_
{zf}^2t^2\right)]\exp\left(-\lambda t\right)\\ 
+ A_{bg},
\label{eqn1:ZF}
\end{split}
\end{equation}
Here $\sigma_{zf}$ is the muon spin relaxation arising from the randomly oriented, static local fields associated with nuclear moments at the muon site. $\lambda$ is associated with the electronic relaxation rate, probably from the Ir. A$_{0}$ and A$_{bg}$ are the initial asymmetry contribution from the sample and background asymmetry which is constant and non-decaying in nature. Solid lines show the fits for both temperatures and clearly suggests the preserved time-reversal symmetry in the superconducting ground state of IrGe within the detection limit of $\mu$SR. 

\begin{figure}[t]
\includegraphics[width=1.0\columnwidth,origin=b]{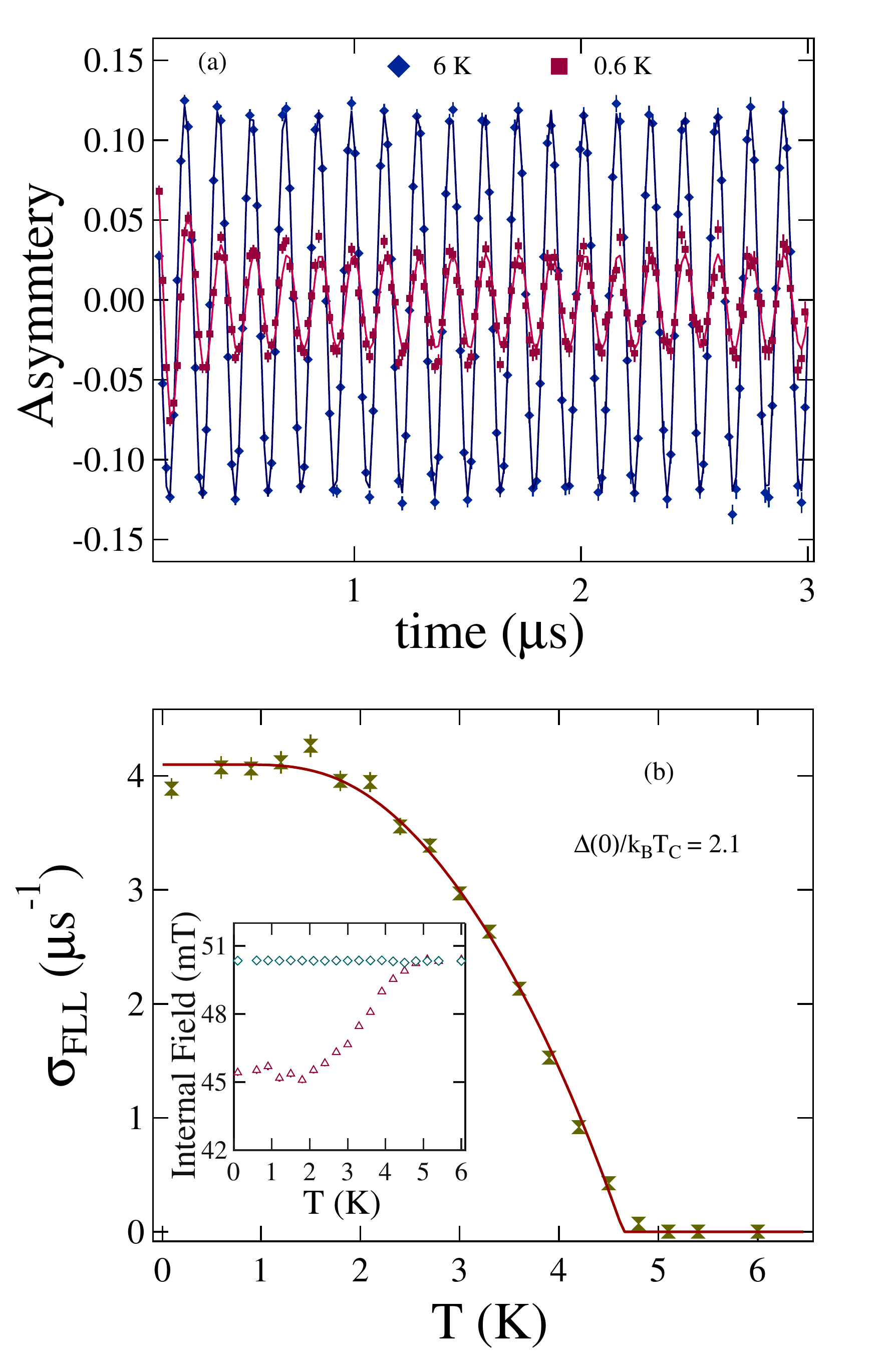}
\caption{\label{Fig6:TF} a) TF-$\mu$SR asymmetry spectra collected above (6.0 K) and below (0.6 K) T$_{C}$ shows different relaxation rates. b) Flux-line lattice contribution to the relaxation rate vs temperature as determined from TF-$\mu$SR spectra where solid line is a fit using s-wave superconducting gap function. Inset: Temperature dependence of internal magnetic field distribution.}
\end{figure}

To explore the nature of the superconducting gap in IrGe at a microscopic level, $\mu$SR measurements in transverse applied fields were carried out. Experiments were performed in field cooled conditions in which a transverse field of 50 mT was applied in the normal state to ensure a formation of well-ordered flux line lattice, and data were collected while warming. The typical polarization spectra in 50 mT collected above, and below T$_{C}$ are presented in \figref{Fig6:TF}(a). The enhanced depolarisation rate below T$_{C}$ reflects the inhomogeneous field distribution of the flux-line lattice (FLL) in the mixed state. The reduced depolarization above T$_{C}$ is due to the randomly oriented static, on the time-scale of the muon, nuclear moments. The TF spectra can be modelled by a sum of N sinusoidal oscillating functions, each damped with a Gaussian relaxation envelope plus a flat background term \cite{TF1,TF2}
\begin{equation}
A (t) = \sum_{i=1}^N A_{i}\exp\left(-\frac{1}{2}\sigma_i^2t^2\right)\cos(\gamma_\mu B_it+\phi)+A_{bg}
\label{eqn1:Tranf}
\end{equation}
A$_{i}$, B$_{i}$, $\sigma_{i}$, $\phi$, and $\gamma_{\mu}$/$2\pi$ = 135.5 MHz/T are initial asymmetry, Ist moment for the ith component of the field distribution, Gaussian relaxation rate, initial phase offset, and muon gyromagnetic ratio, respectively. In this case, N = 2 oscillating functions were sufficient to fit the data where the relaxation from the second component ($\sigma_{2}$) was fixed to zero. It takes into account the non-depolarizing muons which missed the sample and hit the sample holder. The inset of \figref{Fig6:TF}(b) shows the internal field distribution w.r.t. the measured temperature range. B$_{bg}$ corresponds to the background field which is temperature independent, whereas <B> represents the expulsion of magnetic field on entering the superconducting state by reducing the strength of applied field (B$_{app}$) below T$_{C}$ and recovers to B$_{app}$ above T$_{C}$. To calculate the flux-line lattice state contribution to the relaxation rate ($\sigma_{FLL}$), the extra broadening arising from the nuclear moments ($\sigma_{N}$) must be subtracted in quadrature from $\sigma$ using $\sigma_{FLL}$ = $(\sigma^{2}-\sigma_{N}^{2})^{1/2}$. As $\sigma_{N}$ is expected to be temperature independent, its contribution can be taken from the normal state $\sigma$ value. Figure \ref{Fig6:TF}(b) main panel shows the temperature dependence of $\sigma_{FLL}$ and is related to penetration depth via $\lambda^{-2}\propto\sigma_{FLL}$. Considering London local electrodynamics, the temperature dependence of penetration depth for s-wave superconductor in dirty limit (see \secref{Elec}) was modelled using 
\\
\begin{equation}
\frac{\sigma_{FLL}(T)}{\sigma_{FLL}(0)} = \frac{\lambda^{-2}(T)}{\lambda^{-2}(0)} = \frac{\Delta(T)}{\Delta(0)}\mathrm{tanh}\left[\frac{\Delta(T)}{2k_{B}T}\right]
\label{eqn14:swave}
\end{equation}
\\
where  $\Delta$(T) = $\Delta_{0}$ tanh[1.82(1.018($\mathit{(T_{C}/T})$-1))$^{0.51}$] is the BCS approximation for the temperature dependence of the energy gap. The function provides the best fit to the data which is represented by the solid blue line, and yields the superconducting gap value, $\Delta_{0}$/k$_{B}$T$_{C}$ = 2.1 (1). The obtained value is in agreement with the specific heat results and is significantly higher than the predicted BCS value. It indicates a strong coupling nature of the superconducting pairs in IrGe. The presence of strong electron-phonon coupling can be attributed to the low lying phonons arising from the low frequency vibration of the Ir-5d states. The major contribution of Ir-5d states near the Fermi level also gives rise to high SOC (SOC $\propto$ $Z^{4}$). However, nodeless s-wave superconductivity and the presence of TRS is quite surprising and suggests conventional superconductivity in spite of strong SOC and low lying excitation mediating Copper pairing. Further TF measurements on a high quality single crystal are required to exactly determine the role of above mentioned mechanism on the superconducting gap structure.

\begin{table}
\caption{Superconducting and normal state parameters of IrGe determined form various measurements}
\label{Tab:all}
\begingroup
\setlength{\tabcolsep}{22pt}
\begin{tabular}[b]{l c c}\hline\hline
Parameters& unit& IrGe\\
\hline
$T_{C}$& K& 4.74(3)\\             
$H_{C1}(0)$& mT& 13.3(2)\\
$H_{C2}(0)$& T& 1.13(2)\\
$\Delta C_{el}/\gamma_{n}T_{C}$&   &2.7\\
$\theta_{D}$& K& 160(1)\\
$\lambda_{e-ph}$& & 0.78(2)\\[1ex]
$\frac{\Delta^{SH}(0)}{k_{B}T_{C}}$&  &2.3(1)\\[1ex]
$\frac{\Delta^{\mu}(0)}{k_{B}T_{C}}$&  &2.1(1)\\[1ex]
$\xi_{GL}$&  \text{\AA}& 170(1)\\
$\lambda_{GL}$& \text{\AA}& 1750(11)\\
$k_{GL}$& &10(1)\\
$n$&  10$^{28}m^{-3}$& 1.27(2)\\
$m^{*}$&  $m_{e}$& 4.6(4)\\
$l_{e}$&  \text{\AA}& 41(6)\\
$\xi_{0}$&  \text{\AA}& 534(33)\\
$T_{F}$& K& 5088(469)\\[1ex]
\hline\hline
\end{tabular}
\par\medskip\footnotesize
\endgroup
\end{table}

\subsection{Electronic Properties and Uemura Plot}
\label{Elec}
\vspace{-7pt}
In order to confirm the dirty/clean limit superconductivity, the BCS coherence length, $\xi_{0}$ and mean free path $l_{e}$ was determined. Based on the Drude model \cite{Xi_GL}, mean free path can be estimated using the expression $l_{e} = v_{F}\tau$ where $\tau$ is the scattering time and is given by $\tau^{-1}$ = ne$^{2}\rho_{0}$/m$^{*}$. $m^{*}$ is the effective mass of the quasiparticles, and can be evaluated using the following relation: $m^{*}$ = ($\hbar k_{F})^{2}\gamma_{n}$/$\pi^{2}nk_{B}^{2}$ where $k_{F}$ is the Fermi wave vector. Considering a spherical Fermi surface and using the obtained value of
$n$ = 1.27(2) $\times$ 10$^{28}$ m$^{-3}$ from hall measurement, $k_{F}$ = (3$\pi^2n$)$^{1/3}$ was estimated as 0.73(1) $\text{\AA}^{-1}$. After incorporating the obtained values of $k_{F}$, and $n$ together with $\gamma_{n}$ = 1.73 $\times$ 10$^{2}$ Jm$^{-3}$K$^{-2}$, we obtained $m^{*}$ = 4.6(4)m$_{e}$. The Fermi velocity can be determined from the expression $v_{F} = \hbar k_{F}/m^{*}$ = 1.8(1) $\times$ 10$^{5}$ m/s. Using the values of $n$, $m^{*}$, $v_{F}$, and $\rho_{0}$ = 5.57 $\times$ 10$^{-7}$ $\Omega$-m, $l_{e}$ is estimated to be 41(6) $\text{\AA}$. Within the BCS theory, coherence length ($\xi_{0}$) is defined as 0.18$\hbar v_{F}$/$k_{B}T_{C}$ which yields $\xi_{0}$ = 534(33) \text{\AA}. The ratio of $\xi_{0}$/$l_{e}$ = 13(2) indicate the dirty limit superconductivity in IrGe. Using the values of n = 1.27(2) $\times$ 10$^{28}$ m$^{-3}$ and m$^{*}$ = 4.6(4)m$_{e}$, the effective Fermi temperature of a 3D system can be calculated by the following relation \cite{TF}: $k_{B}T_{F}$ = $(3\hbar^{3}\pi^{2}n)^{2/3}/{2m^{*}}$ which came out to be 5088(469) K.

\begin{figure} 
\includegraphics[width=1.0\columnwidth, origin=b]{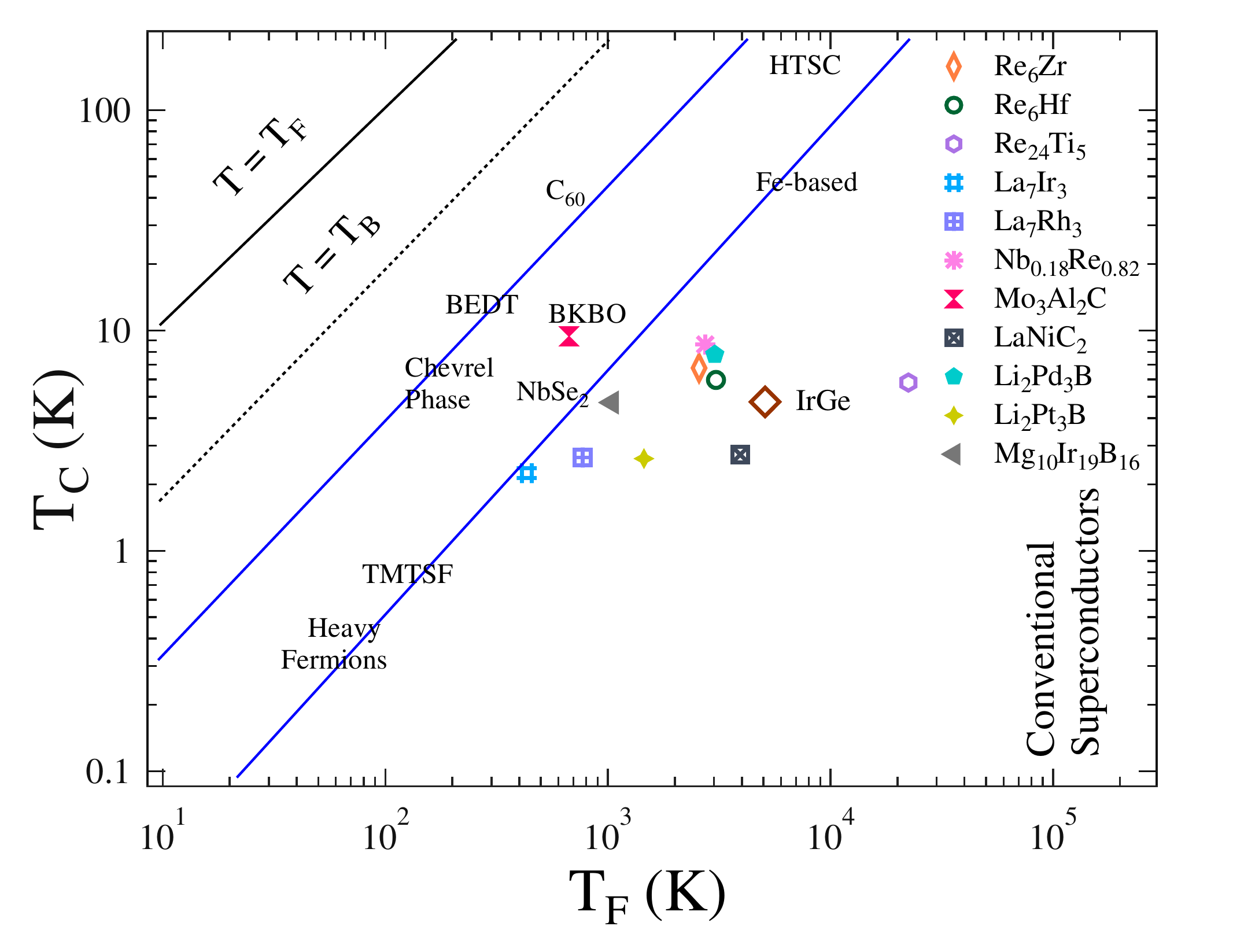}
\caption{\label{Fig7:Uemura} Uemura plot representing transition temperature versus Fermi temperature for various superconductors belong to different classes where the region between two solid blue lines represent the band of unconventionality. IrGe is represented by a hollow red marker.}
\end{figure}

Superconductors can be classified as conventional/unconventional based on the ratio of T$_{C}$/T$_{F}$ provided by Uemura \textit{et al.} \cite{U1,U2,U3,U4,U5}. Compounds belonging from different classes such as heavy fermions, Chevrel phases, high T$_{C}$, and Fe-based superconductors fall in the category of unconventional as their ratio lies in 0.01 $\leq$ T$_{C}$/T$_{F}$ $\leq$ 0.1 range. The obtained value of T$_{C}$/T$_{F}$ = 0.0009 for IrGe lies significantly outside the range of unconventional superconductors as is shown by a hollow red marker in \figref{Fig7:Uemura}. The result is surprising as IrGe falls in the strong-coupling regime which cannot be explained by the BCS theory yet it is placed away from the band of unconventionality.

In summary, a detailed examination of the physical properties of IrGe is reported using various techniques which confirmed the bulk nature of superconductivity by exhibiting a T$_{C}$ at 4.74(3) K. All the obtained parameters related to the superconducting and normal state are listed in \tableref{Tab:all}. Zero field specific heat and TF-$\mu$SR study reveal an isotropic nodeless superconductivity in strong coupling regime. The results are in contrast to other low excitations mediated strongly coupled superconductors where anisotropic, multigap, and multiband type superconducting gap structure were observed. Zero-field $\mu$SR provide the evidence of absence of the time reversal symmetry breaking in IrGe. Therefore, from the microscopic measurements, the rattling mediated superconductivity in IrGe seems to be of conventional type. However, the strong coupling nature originated from the rattling modes still seeks further examination through a detailed theoretical investigation of the Fermi surface and single crystal measurements.

\section{Acknowledgments} Arushi acknowledges the funding agency, University Grant Commission (UGC) of the Government of India for providing SRF fellowship. R.~P.~S.\ acknowledges Science and Engineering Research Board, Government of India for the Core Research Grant CRG/2019/001028. Department of Science and Technology, India (Grant No. SR/NM/Z-07/2015) for the financial support and Jawaharlal Nehru Centre for Advanced Scientific Research (JNCASR) for managing the project. We thank ISIS, STFC, UK for the beamtime to conduct the $\mu$SR experiments \cite{MuData}.

\end{document}